# The role of atomic orbitals of doped earth-abundant metals on designed copper catalytic surfaces

Dequan Xiao[a,*] and Trevor Callahan[a]

a. HIGA Polymer Materials Laboratory, Center for Integrative Materials Discovery, Department of Chemistry and Chemical Engineering, University of New Haven, West Haven, CT 06516

* corresponding author: dxiao@newhaven.edu

**Abstract** It is a general challenge to design highly active or selective earth-abundant metals for catalytic hydrogenation. Here, we demonstrated an effective computational approach based on inverse molecular design theory to deterministically search for optimal binding sites on Cu (100) surface through the doping of Fe and/or Zn, and a stable Zn-doped Cu (100) surface was found with minimal binding energy to H-atoms. We analyze the electronic structure cause of the optimal binding sites using a new quantum chemistry method called orbital-specific binding energy analysis. Compared to the 3d-orbitals of surface Cu atoms, the 3d-orbitals of surface Zn-atoms show less binding energy contribution and participation, and are much less influenced by the electronic couplings of the media Cu atoms. Our study provides valuable green chemistry insights on designing catalysts using earth-abundant metals, and may lead to the development of novel Cu-based earth-abundant alloys for important catalytic hydrogenation applications such as lignin degradation or $CO_2$ transformation.

## Introduction

Designing highly active or selective catalysts using earth-abundant metals is consistent with one of the principles of green chemistry that is to design greener and safer chemicals.[1] A few experimental[2-7] and computational[8,9] design efforts have been made to develop novel heterogeneous Cu-based catalysts by doping with other metals, for various interesting applications such as the hydrogenation of lignin[2,3], syngas[4,5], $CO_2$[10], and biomass derivatives[6,7].

However, it remains a general challenge to design highly active or selective earth-abundant alloy catalysts through deterministic search. In this work, we will demonstrate an effective way to computationally search for optimal catalytic active sites on Cu (100) surface through the doping of earth-abundant metals using the inverse molecular design method[11], and then reveal the role of atomic orbitals of the effective doped metals using a new quantum chemistry method called orbital-specific binding energy analysis. Our work will provide valuable green chemistry insights on designing catalysts using earth-abundant metals, which can lead to the development of novel earth-abundant alloy catalysts for important applications such as lignin degradation and $CO_2$ transformation. For example, the designed optimal surface binding sites here may be created by the fabrication techniques used for single-atom catalysis[12-14] or by the search of potential bulk alloys in practice.

Doping Cu with other metals (either noble or earth-abundant metals) has shown to be an effective way in practice to generate promising hydrogenation catalysts. For examples, Cu/ZnO/Al$_2$O$_3$ is an industrial catalyst for methanol synthesis through the hydrogenation of CO[4], and was also used for water-gas shift reaction at relatively low temperature[5]. Cu-Zn alloy was used to synthesize ethylene glycol from the hydrogenation of dimethyl oxalate with remarkable activity.[15] A bimetallic Cu-Pd alloy was placed on the support of reduced graphene oxide showing improved catalytic activity.[7] Recently, a Cu-doped hydrotalcite show high yields of liquid fuels and aromatics from the hydrogenation of lignin under relatively mild condition,[2,3] was later found to have high selectivity on carbonyl groups or double-bonds at relatively low temperature.[16]

We aim to minimize the binding affinity to H-atoms on the Cu-based catalytic surfaces, which can lead to the lowering of activation energy barrier for catalytic hydrogenation. H-adsorption energy is correlated to catalytic activities as "volcano" plots[17]. Strong hydrogen binding may increase the concentration of H-atoms adsorbed on catalytic surfaces; however, it may also increase H-atom release barrier for the next hydrogenation step. Weak hydrogen binding on surfaces facilitates the release of H-atoms; however, it may also decrease the concentration of adsorbed H-atoms on catalytic surfaces. This is known as Sabatier principle.[18] Here, we will minimize the H-release barrier on catalytic surfaces while maintains the favorable binding of H-atoms on the surfaces, which can lead to highly active hydrogenation catalysts that work under low temperature.

We will use the inverse molecular design methods in tight-binding framework to optimize the Cu (100) surface. The motivation of inverse molecular design is to search for optimal molecular or chemical structures guided by efficient optimization algorithms.[19] Early studies of inverse design methodologies focused on finding optimum bandgap atomic configurations using the effective inverse band-structure (IBS) approach, based on the simulated annealing algorithm[20] or genetic algorithm.[21] In 2006, a new approach based on linear combination of atomic potential (LCAP) was developed at the level of density functional theory (DFT),[22] for optimizing polarizabilities and hyperpolarizabilities. In 2008, Xiao et al. implemented the LCAP approach into the tight-binding (TB) framework using a Hückel-type Hamiltonian that is shown to be effective at locating optimal structures within a large number ($10^2$–$10^{16}$) of possible chemical structures.[23] Later, Xiao et al. developed the extended-Hückel (EH) generalization of the Hückel TB-LCAP approach for the purpose of optimizing molecular sensitizers for dye-sensitizer solar cell applications.[24] Herein, we will adapt the TB-LCAP inverse molecular design approach to search for optimal binding sites to H-atoms through the doping of earth-abundant metals (i.e., Fe, and Zn) on Cu (100) surfaces.

We will analyze the role of atomic orbitals of doped earth-abundant metals on Cu (100) surfaces using a new quantum chemistry analysis method called orbital-specific binding energy analysis. Revealing the role of specific atomic orbitals of surface atoms at active sites upon chemisorption of H-atoms will help us to understand the electronic structure cause of the optimal active site, leading to new insights (thus innovative approaches) to design highly active/selective earth-abundant catalysts. The influence of atomic orbitals to the binding of H-atoms has been studied through the analysis of density of states (DOS) of bulk alloys before and after H-atoms binding.[25] However, this analysis can only use the change in DOS or projected DOS (e.g., DOS of d-bands) to imply the interaction of surface metallic orbitals with the adsorbed H-atoms, and there is no quantitative analysis of the binding energy contribution per atomic orbital or electronic band. To solve this problem, a d-band model was successfully developed by Hammer and Nørskov [26] by decomposing the overall adsorbate-surface binding energy into the contributions from sp bands and d-band based using a few fitting parameters, where the d-band is treated as a single d-electronic level with energy $\varepsilon_d$ and one coupling matrix element V between d-band and bonding or anti-bonding orbital of the adsorbate. The d-band model was successfully used to design novel heterogeneous catalysts. Here, we developed a computational analysis method based on the analytic derivation of the electronic structure theory without additional fitting parameters. We will provide new insights for the binding energy contributions for the overall binding energy from the viewpoint of individual atomic orbitals, instead of the viewpoint of electronic bands.

In the orbital-specific binding energy analysis, the overall H-atom binding energy is decomposed into the binding energy contributions of all (e.g., hundreds of) the atomic orbitals of the metallic atoms, without additional approximation or fitting parameters. In addition, we will use Green's function methods to analyze the influence of media atoms to the surface atomic orbitals that are in direct contact with H-atoms. The Green's function methods have been previously used to study the atomic orbital couplings between donor-and-acceptor for electron transfer study,[27] and electron transport for molecules or solid-state materials[28]. Using the Green's function method in the orbital-specific binding energy analysis here, we can tell the role of earth-abundant metal atomic orbitals upon the chemisorption of H-atoms.

In the followings, we first describe the theory of inverse molecular design in tight-binding framework for H-atom binding energy minimization, and also the theory for orbital-specific binding-energy analysis. We will then show the TB-LCAP search for optimal surface binding sites through the doping of Zn and/or Fe atoms on the Cu (100) surfaces. After that, we will verify the accessibility of such surface and its optimal H-binding property using geometry relaxation at the density functional theory level in the framework of infinitely repeated lattice units. We will analyze the change of density of states upon the binding of H-atoms for pure Cu (100) surfaces and the designed Zn-doped Cu (100) surfaces, respectively. Lastly, we will analyze the role of surface atomic orbitals by performing the orbital-specific binding energy analysis for H-atom adsorption on pure Cu (100) surface, and the Zn-doped Cu (100) surface, respectively.

## Theories

**Minimization of the H-atom binding energy by the Inverse molecular design approach**

The following describes the extended-Hückel tight-binding linear combination of atomic potential (TB-LCAP) method as implemented for optimizing H-atom binding energy on the Cu (100) surfaces.

The extended Hückel approach[29,30] has been widely used for electronic structure and materials property calculations for molecules and solid-state structures.[31] The time-independent Schrödinger equation in the extended Hückel framework can be represented in matrix form,

$$HC = ESC \quad (1)$$

where $H$ is the extended Hückel Hamiltonian in the basis of Slater-type atomic orbitals (AO's), $C$ is the matrix of eigenvectors, $E$ is the diagonal matrix of eigenstate energies, and $S$ is the overlap matrix of the Slater-type atomic orbitals.

In the extended Hückel TB-LCAP formula,[24] participation coefficients $b_i^{(A)}$ are introduced to represent the occurrence probability of $N_{type}^i$ optional atom-type A at site i, with the conditions of $\sum_A b_i^{(A)} = 1$ and $0 \leq b_i^A \leq 1$.

The diagonal matrix elements of the EH/TB-LCAP Hamiltonian are related to the participation coefficients $b_i^{(A)}$ by

$$H_{i\alpha,i\alpha} = \sum_{A=1}^{N_{type}^i} b_i^{(A)} h_{i\alpha,i\alpha}^{(A)} \qquad (2)$$

where $\alpha$ is the index for atomic orbitals. Note that for the special case of $b_i^{(A)} = 1$, $H_{i\alpha,i\alpha} = h_{i\alpha,i\alpha}^{(A)}$, the site energy of AO $\alpha$ of atom type $A$. The off-diagonal matrix elements are related to the participation coefficients $b_i^{(A)}$ by

$$H_{i\alpha,i\beta} = \sum_{A=1}^{N_{type}^i} \sum_{A'=1}^{N_{type}^j} b_i^{(A)} b_j^{(A')} h_{i\alpha,j\beta}^{(A,A')} \qquad (3)$$

Here, $\beta$ is the index of atomic orbitals, $h_{i\alpha,j\beta}^{(A,A')}$ is the original extended Hückel AO interaction energy between atomic orbitals $\alpha$ (of atom type $A$ at site $i$) and $\beta$ (of atom type $A'$ at site $j$). [24]

The overlap matrix (S) element is computed by

$$S_{i\alpha,i\beta} = \sum_{A=1}^{N_{type}^i} \sum_{A'=1}^{N_{type}^j} b_i^{(A)} b_j^{(A')} S_{i\alpha,j\beta}^{(A,A')} \qquad (4)$$

where $S_{i\alpha,j\beta}^{(A,A')}$ is the overlap matrix element between $\alpha$ (of $A$ at site $i$) and $\beta$ (of $A'$ at site $j$).

The participation coefficients $b_i^{(A)}$ are initialized randomly, and subsequently optimized for the target property, i.e., H-atom binding energy here. The chosen Slater-type atomic orbitals include 3d, 4s and 4p atomic orbitals for Cu, Fe, Zn, and 1s orbital for H.

The goal of TB-LCAP search here to minimize is the magnitude of binding energy of H-atoms on the catalytic surface.

$$E_{binding} = E_{Cu-H} - (E_{Cu} + E_H) \qquad (5)$$

Here, $E_{Cu-H}$ denotes the electronic energy of the Cu (100) surface bound by H-atoms, $E_{Cu}$ is the energy of Cu (100) surface without the binding of H-atoms, and $E_H$ is the energy of free H-atoms.

$E_{binding}$ is minimized (in terms of magnitude) with respect to the participation coefficients $b_i^{(A)}$ using standard optimization techniques (e.g., the quasi-Newton BFGS algorithm), computing the gradients of $E_{binding}$ in terms of $b_i^{(A)}$ using standard finite difference expressions.

## Computational methods

The accessibility and optimal property of the designed surfaces were verified by advanced DFT calculations. Geometry relaxation of the pure Cu or Zn-doped Cu lattice was performed by the DFT (PW91/GGA) method with periodic boundary condition using a plane-wave basis set and the ultrasoft Vanderbilt pseudopotential approximation, as implemented in the Vienna ab initio Simulation Package (VASP/VAMP).[32] A lattice unit with 32 metallic atoms was used to simulate the pure Cu or Zn-doped Cu lattice. The wave function cutoff was 400 eV, and single γ-point sampling was used when surface geometry was performed. For the projected density of states analysis using VASP, we used Monkhorst Pack 9x9x1 for the number of k-points.

The following is the procedure for computing the H-binding energy on Zn-doped (or pure) Cu (100) surfaces using the DFT calculations with periodic boundary condition. We first relaxed the lattice parameters of the Zn-doped Cu lattice unit. After that, we computed the energy of the Zn-doped Cu surface ($E_{Cu}$ in equation 5) without the adsorbed H-atoms. Then, we attached the H-atom and relaxed only the surface atoms by fixing the atomic configuration of the bottom two layers. In this way, we obtained the total energy of the surface ($E_{Cu-H}$ in equation 5) after the binding of H-atoms.

## Results and discussion

### 1. TB-LCAP search for optimal binding sites through atomic doping on Cu (100) surfaces

As shown in Figure 1, we first define a search framework based on the copper (100) surface. On this surface, a hydrogen atom is placed in the four-fold hollow binding site, as it is the most likely binding sites on the Cu (100) surface. We allow four atom sites to vary with the choice of Cu, Fe, and Zn atom types, where the four atoms sites are in direct-contact with the H-atom as highlighted in blue in Figure 1a. We then run the TB-LCAP inverse molecular design program with 10 random initiations. A typical search path that leads to the lowest binding energy is shown Figure 1c. The search framework starts at a random structure with a binding energy of -3.2 eV, and then vary deterministically to a structure with a binding energy of -2.5 eV. The optimized active site has three Zn atoms, in an atomic configuration shown in Figure 1d. Here, we found that gradient of H-atom binding energy hypersurface versus the alchemical structure variation is well-defined, providing gradients to the final structure.

### 2. Analyzing accessibility and validity of the search results on the surfaces of periodic units using DFT calculations

Based on the searched optimal binding site, we constructed a minimal repeat lattice unit containing the whole optimal binding site (see Figure S1). We performed the surface geometry relaxation of the lattice repeat unit, using the DFT calculations with periodic boundary condition. The atom configuration of the relaxed surface-binding site remains to be stable, suggesting the accessibility of such binding site in experiments. Based on the relaxed geometry, we computed the binding energy of H atoms on the fourfold hollow-site (site 2 in Figure 2) under periodic boundary condition, which is -0.427 eV. Similarly, we calculated the binding energy of H-atoms on the four-fold hollow-site (site 8 in Figure 2) on pure Cu (100) surface using the same-size lattice under periodic boundary condition, and the calculated binding energy is -0.814 eV. Therefore, we verify that the Zn-doped Cu binding site indeed has less binding energy than the Cu fourfold hollow sites on periodic solid-state surfaces, indicating the effective search by the TB-LCAP inverse molecular design method.

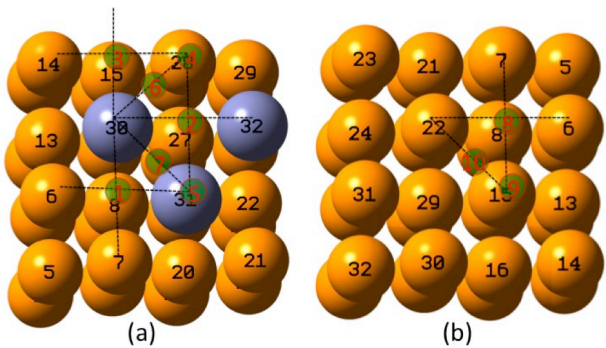

**Figure 2.** Potential binding sites 1-10 (red) on (a) Zn-doped and (b) pure Cu (100) surfaces for H-atoms.

The doping of Zn atoms to the Cu (100) surfaces also creates other new potential H-atom binding sites (see sites 1-7 in Figure 2). So, we computed the H-atom binding energies for all these sites, and compared them to the possible binding sites (see sites 8-10, Figure 2) on pure Cu (100) surfaces. The calculated binding energies are listed in Table 1. We found that the highest possible binding energy (-0.702 eV) on the Zn-doped Cu (100) surface is actually lower than the highest possible binding (-0.814 eV) on pure Cu (100) surface. Thus, the entire Zn-doped Cu (100) surface is expected to have lower overall binding affinity to H-atoms than pure Cu (100) surface, indicating a meaningful design of optimal catalytic surfaces by the TB-LCAP search. The finding of low H-atom binding affinity of Zn-doped Cu surface may be related to the reported high-activity of Zn-doped Cu catalysts in the literature.[4,5,15]

For better understanding the electronic structure cause of lowering H-atom binding energy through the doping of Zn-atoms, we will perform the following quantum chemistry analyses based on the found fourfold hollow site (site 2 in Figure 2) on Zn-doped Cu (100) surface and compare it to the fourfold hollow site on pure Cu (100) surface.

**Table 1.** Binding energy (eV) of H-atoms in different positions on the pure and Zn-doped Cu (100) surfaces in Figure 2.

| Site No. | Site type | Binding energy (eV) |
|---|---|---|
| 1 | fourfold hollow | -0.655 |
| 2 | fourfold hollow | -0.427 |
| 3 | fourfold hollow | -0.702 |
| 4 | top | -0.175 |
| 5 | top | -0.328 |
| 6 | bridge | -0.490 |
| 7 | bridge | -0.328 |
| 8 | fourfold hollow | -0.814 |
| 9 | top | -0.140 |
| 10 | bridge | -0.664 |

### 3. Density of states analysis for the binding of H-atoms

We performed the density of state (DOS) analysis for pure and Zn-doped Cu (100) surfaces, at the DFT level with periodic boundary condition. As shown in Figure 3, there are changes on projected DOS for all the valence orbitals (i.e., 3d, 4s, and 4p) of Cu and Zn upon the binding of H-atoms. For example, the Cu 4s-orbital shows clear change (see Figures 3a and 3d). In Figure 3f, the project DOS of Cu 3d-orbitals on the Zn-doped Cu (100) surface shows a spread-out upon the binding of H-atoms. In Figure 3i, there is an apparent energy shift of the Zn 3d-orbitals upon the binding of H-atoms. From the projected DOS analysis, we can observe changes of projected orbital density or energy shifts of atomic orbitals. However, we do not know qualitatively what are the binding energy contributions by specific atomic orbitals. In the following, we will perform the orbital specific binding energy analysis for the pure Cu and Zn-doped (100) surfaces, respectively.

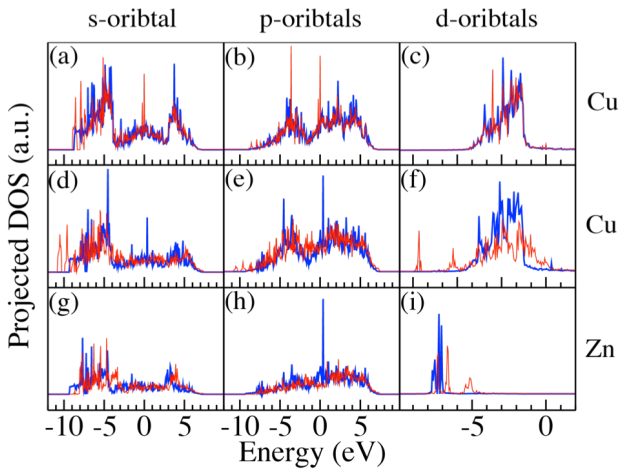

**Figure 3.** Projected density of states (DOS) of 4s, 4p, and 3d orbitals for Cu-atoms on pure Cu (100) surface (a-c), Cu-atoms on Zn-doped Cu (100) surface (d-f), and Zn-atoms on Zn-doped Cu (100) surface (g-i). The blue curves are the DOS without adsorbed H-atoms, and the red curves are the DOS with adsorbed H-atoms on the surfaces.

## 4. Orbital-specific binding energy analysis for H-atom adsorption

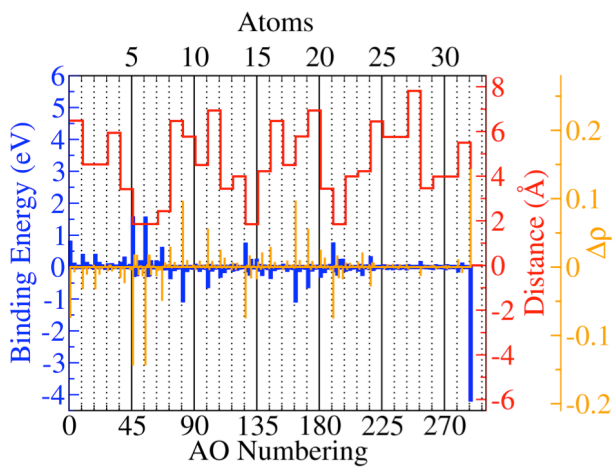

**Figure 4.** (a) Binding energy contributions by individual atomic orbitals: 1-289 on (b) pure Cu (100) surface. For Cu atoms, the AO orbitals are numbered in the order of **4s**, **4$p_x$**, **4$p_z$**, **4$p_y$**, **3$d_{x^2-y^2}$**, **3$d_{z^2}$**, **3$d_{xy}$**, **3$d_{xz}$**, and **3$d_{yz}$**. AO 289 is the 1s orbital of hydrogen atom. The red curve shows the distance of the AO center to the adsorbed H-atom, and the orange lines denote the change of electron population on individual AO upon the binding of H-atom. Positive Δρ indicates an increase of electron population, and negative Δρ indicates a decrease of electron population.

We performed the orbital-specific binding energy analysis at the level of the extended Hückel tight-binding theory. We used the lattice repeat units optimized at the DFT level as the crystal slab models to perform the orbital-specific binding energy analysis. By the extended Hückel calculations, the Zn-doped Cu slab (28 Cu atoms and 3 Zn atoms) shows lower binding energy (-2.64 eV) than the pure Cu slab (-3.29 eV), which is consistent with the DFT calculation with periodic boundary condition. Thus, the binding energy analysis based on extended Hückel tight-binding calculations and the crystal slab models is appropriate for us to understand the role of atomic orbitals upon the binding of H-atoms.

As shown in Figure 4, upon the binding of H-atoms on pure-Cu (100) surface, it is interesting to find that electron population change on the AO is closely correlated to the nature of binding energy contribution: an increase of electron population (positive Δρ) on an atomic orbital causes a negative (favored) contribution to the binding energy; while a decrease of electron population (negative Δρ) on an atomic orbital causes a position (disfavored) contribution to the binding energy. On the adsorbed H-atom, there is net gain of electron population after binding to the four-fold hollow site, indicating electron transfer from the Cu (100) surface to the adsorbed H-atom. The increase of electron population on H-atom 1s-orbital causes a negative contribution to the binding energy. Overall, hydrogen 1s-orbital contributes a negative binding energy (-4.0 eV). We find that some atomic orbitals (particularly those on Cu-atoms in direct-contact with the H-atom) have apparent positive binding energy contributions, e.g., Cu6 4s-orbitals (+1.52 eV), and Cu7 4s-orbital (+1.52 eV). Some atom orbitals contribute negatively to the binding energy, e.g., Cu10 4s-orbital (-1.04 eV) and Cu19 4s-orbital (-1.06 eV).

Overall, the atomic orbitals on pure Cu (100) surface have a net positive binding energy contribution. The Cu atomic orbitals in direct-contact with the H 1s-orbital show large positive contributions. The atomic orbitals distant to the H-atom tend to show minor contributions to the binding energy. But, some distant Cu-atom orbitals can still show relative large binding energy contributions such as Cu1 4s-orbital.

The binding energy contribution for each atom in the pure Cu slab can be found in Figure S2. The five direct contact atoms Cu6, Cu7, Cu8, Cu15, and Cu22 make positive contributions to the binding energy. The rest indirect-contact Cu atoms contribute positively or negatively to the overall binding energy. The H-atom shows the largest binding participation percentage (60%), followed by direct-contact atoms Cu6 (10%) and Cu7 (10%).

Figure 5 shows the orbital specific binding analysis for the Zn-doped Cu slab. Again, we found that positive Δρ is correlated to a negative contribution to the binding energy; negative Δρ is correlated to a position contribution to the binding energy. Upon the binding of the H-atom, the Zn-doped

Cu binding site shows slightly higher gain of electron population on the H-atom 1s-oribtal than that of the pure Cu (100) surface. Meanwhile, a much higher loss of electron population is found on Cu27 atomic orbitals than any the atomic orbitals of any other Cu atom on the pure Cu slab. For Zn atoms, Zn30 4s and Zn31 4s orbitals show strong positive contribution to the overall binding energy.

Overall, as shown in Figure S3, the H-atom contributes a negative binding energy (-5.0 eV). For the direct-contact Cu-atoms, Cu28 shows a positive binding energy contribution, and Cu27 shows a negative binding energy contribution. All three Zn-atoms show positive binding energy contributions. The adsorbed H-atom shows the highest binding participation (60%), followed by Cu28 (18%) and Zn30 (5%). The indirect-contact atom Cu1 shows a moderately high binding participation (4%).

## 5. The influence of media atoms orbitals

As shown in Figures 4 and 5, the media atoms actively participate in the binding of H-atom. If we assume that the H 1s-orbital is localized (a valid approximation), the media atom orbitals can influence the H-binding through the atomic orbitals of Cu atoms that are in direct-contact with the adsorbed H-atom.

Figure 6a (top panel) shows the orbital-specific binding energy contribution of the core atoms on Zn-doped copper

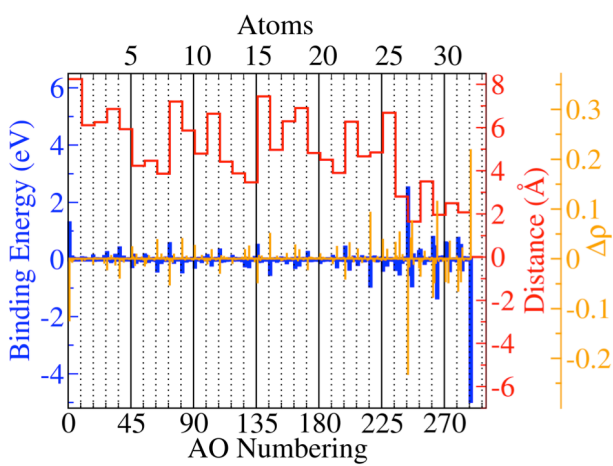

**Figure 5.** (a) Binding energy contributions by individual atomic orbitals: 1-289 on (b) Zn-doped Cu (100) surface. For Cu or Zn atoms, the AO orbitals are numbered in the order of **$4s$, $4p_x$ $4p_z$, $4p_y$, $3d_{x^2-y^2}$, $3d_{z^2}$, $3d_{xy}$, $3d_{xz}$**, and **$3d_{yz}$**. AO 289 is the 1s orbital of hydrogen atom. The red curve shows the distance of the AO center to the adsorbed H-atom, and the orange lines denotes the change of electron population on individual AO upon the binding of H-atom. Positive Δρ indicates an increase of electron population, and negative Δρ indicates a decrease of electron population.

(100) surface, without the presence of any media atoms. We found that H 1s orbital has the largest binding energy contribution, followed by Cu2 4s-orbital. Figure 6a (bottom panel) shows the orbital-specific binding energy contribution of the core atoms on Zn-doped copper (100) surface, with the coupling of all the media atoms. We found very strong binding energy contribution from Cu2 4s-orbital (+50.08 eV), $3d_{x^2-y^2}$-orbital (-56.5eV), and $3d_{yz}$-orbital (-67.2eV). Zn5 4s-ortial shows large binding energy contribution (+20.0 eV). Indeed, in Figure S4a (bottom panel), atom Cu2 shows the highest binding participation percentage (61%), with a large positive contribution to the overall binding energy. Other direct-contact atoms (e.g., Zn) show moderate binding energy contributions, and weak low binding participation percentage. After the coupling of media atoms, the binding participation of the H-atom decreases to 10%. The overall binding energy calculated by the Green's function method is -2.64eV for the Zn-doped Cu slab, reproducing the binding energy calculation result based on all atoms.

In contrast, for the pure Cu (100) lab without the presence of media atoms (see Figure 6b, top panel), Cu1 4s-orbital shows the strongest binding energy contribution, while the other core Cu atoms show similar strong AO binding energy contributions. Atoms Cu2-5 show strong binding energy contributions from the 4s-orbitals. Meanwhile, Cu 3d-orbitals also show strong binding energy contributions, e.g., $3d_{xz}$ orbitals of Cu2 and Cu4, and $3d_{yz}$-orbitals of Cu3 and Cu5.

After the coupling of the media Cu atoms (Figure 6b, bottom panel), all of the core Cu 4s-orbitals show strong binding energy contributions, with stronger contributions from Cu2 and Cu5 than those from the other three core Cu-atoms. Overall, the H atom shows a 35% of binding participation, with a negative contribution (-35.0 eV) to the overall binding energy. The five core Cu atoms show similar binding participation percentages: Cu1: 12%, Cu2: 18%, Cu3: 8%, Cu4: 7%, and Cu5: 20%. The overall binding energy by the Green's function method is -3.29eV for the pure Cu slab, reproducing the binding energy calculation result based on all of the atoms.

By the analysis in Figure 6, we notice the difference in binding energy contributions of the 3d-orbitals between core Cu and Zn atoms. Without the media atom coupling, Cu 3d-oribtals show strong binding energy contributions on the pure Cu binding site. However, Zn 3d-orbitals do not show strong binding energy contributions on the Zn-doped Cu binding site. After the coupling by the media atoms, at the Zn-doped Cu binding site, only Cu2 3d-oribtals show strong binding energy contribution. Zn 3d-orbitals do not shows strong binding energy contribution, indicating the media Cu-atoms have very week coupling to the Zn 3d-orbitals. However, at the pure Cu binding sites, all the core Cu 3d-orbitals show strong binding energy contribution and binding participation, indicating the mediate Cu atoms has strong coupling to the core Cu-atoms. In addition, we clearly observe in Figure 6 that the 4s orbitals of Cu and Zn show strong participation in the binding energy, which is consistent with the results of project DOS analysis at the DFT level with periodic boundary condition.

## Conclusions

We demonstrated an effective design of optimal H-atom binding sites on Cu (100) surface by doping Zn or Fe atoms using TB-LCAP inverse molecular design methods, and found a four-fold hollow site with 2 Cu atoms and 3 Zn atoms with lower binding energy to H-atoms than pure Cu (100) surface. We verified the accessibility and optimal properties of the Zn-doped Cu (100) surface using density functional theory with periodic boundary condition. Project DOS analysis indicates the 3d, 4s, and 4p orbitals of Cu and Zn actively participate in the binding of H-atoms.

We provided new insights on understanding the role of atomic orbitals upon the binding of H-atoms on the pure or Zn-doped Cu (100) surfaces through a new quantum chemistry method called the orbital-specific binding energy analysis. First, we found that the electron population change (Δρ) per atomic orbital is closely correlated to the AO's binding energy contribution: positive Δρ usually causes negative-value (favored) contribution; and negative Δρ usually causes positive-value (disfavored) contribution. Second, the net binding energy contribution of H-atoms is negative (favored), and positive (disfavored) by the pure or Zn-doped Cu (100) surface. The core atomic orbitals usually show stronger binding energy contributions and thus high binding participation than the media atomic orbitals. Even though the mediate atoms show minor binding contribution individually, they collectively show important influence to the over binding energy.

The effectiveness of lowing of H-atom affinity by the doping of Zn on Cu (100) surface can be understood by the distinguished atomic orbitals binding energy contributions between Zn and Cu atoms. On the Zn-doped Cu (100) surface, the binding participation is dominated by one of the core Cu-atoms (Cu28, see Figure S2), and the three Zn-atoms show weak binding participations. The Zn atomic orbitals have less binding participation than Cu atomic orbitals, without the presence of media atoms. After the coupling of media atoms, the binding energy contributions by the Zn atomic orbitals (particularly 3d orbitals) are less influenced compared to the core Cu atomic orbitals. In general, we found that Zn atomic orbitals show less participation in the binding of H-atoms than Cu atomic orbitals.

In summary, we designed an optimal catalytic binding surfaces through the doping of Zn atoms on Cu (100) surfacess, and we explored the effectiveness of the doping of earth-abundant metal Zn through a thourough analysis of the role of metallic atomic orbitals upon the binding of H-atoms. Our study provides valuable green chemistry insights on designing catalysts using earth-abundant metals, and may lead to the development of novel Cu-based earth-abundant alloys for important catalytic hydrogenation applications such as lignin degradation or $CO_2$ transformation.

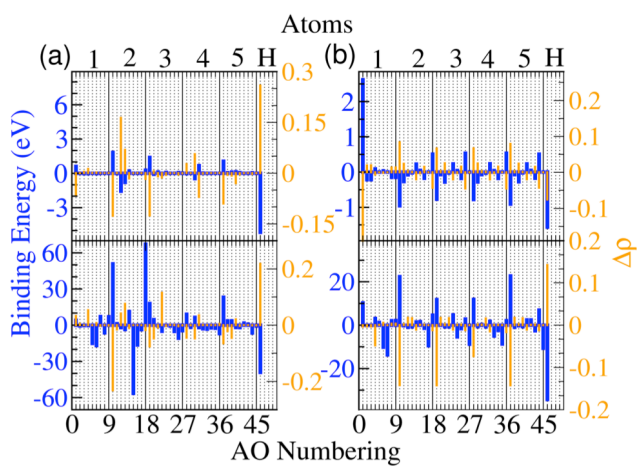

**Figure 6.** Binding energy contributions by individual atomic orbitals (1-46) of the core atoms in (a) the Zn-doped Cu (100) slab and (b) the pure Cu (100) slab. For Cu or Zn atom, the AO orbitals are numbered in the order of **$4s$**, **$4p_x$**, **$4p_z$**, **$4p_y$**, **$3d_{x^2-y^2}$**, **$3d_{z^2}$**, **$3d_{xy}$**, **$3d_{xz}$**, and **$3d_{yz}$**. AO 46 is H 1s-orbital. The top panels indicate the binding energy contributions (blue lines) of AOs without the presence of the media Cu-atoms; and the bottom panels indicate the binding energy contributions (blue lines) of AOs, with the presence of the media Cu-atoms. The orange lines denote the change of electron population on individual AO upon the binding of H-atom. Positive Δρ indicates an increase of electron population, and negative Δρ indicates a decrease of electron population.

## Acknowledgements

This work was supported by the new faculty startup fund and University Research Scholar award to DX provided by the University of New Haven, and the research fund support provided by Higasket Plastics Group Co. Ltd. DX acknowledges supercomputer time from Open Science Grid.